\documentclass[12pt,a4paper]{article}
\usepackage{epsfig}
\begin{document}
\title{ Non-universal gauge boson $Z'$ and the spin correlation of top quark
pair production at $e^{-}e^{+}$ colliders}
\author{Chong-Xing Yue and  Li-Na Wang\\
{\small  Department of Physics, Liaoning Normal University, Dalian
116029, China}\thanks{E-mail:cxyue@lnnu.edu.cn}\\}
\date{\today}

\maketitle
\begin{abstract}
 In the off-diagonal basis, we discuss the contributions
of the non-universal gauge boson $Z'$ predicted by the
topcolor-assisted technicolor ($TC2$) model to the spin
configurations and the spin correlation observable of the top quark
pair production via the process $e^{-}e^{+}\rightarrow t\bar{t}$.
Our numerical results show that the production cross sections for
the like-spin states, which vanish in the standard model, can be
significantly large as $M_{Z'}\approx \sqrt{S}$. With reasonable
values of the $Z'$ mass $M_{Z'}$ and the coupling parameter $k_{1}$,
 $Z'$ exchange can generate
large corrections to the spin correlation observable.

\vspace{1cm}


\end{abstract}

\newpage
\noindent{\bf I. Introduction}

Although the standard model $(SM)$ has been precisely tested by
experiments, it is widely believed to be only an effective theory
valid below some high energy scales. New physics should exist at
energy scales around $TeV$. Most of the new physics models predict
the existence of the new neutral gauge boson $Z'$ associated with
the extra $U(1)$ subgroup of the underlying group. The
phenomenology of this kind of new particles has been extensively
studied in either model-dependent or model-independent
approaches[1], whose presence will be of great interest in the
next generation collider experiments [2].

Topcolor scenario is attractive because it explains the large top
quark mass and provides possible dynamics mechanism of electroweak
symmetry breaking $(EWSB)$[3]. A common feature of this kind of
models is that the $SM$ gauge groups are extended at energy well
above the weak scale. Breaking of the extended gauge groups to their
diagonal subgroups produces new massive gauge bosons. To generate
enough large mass for top quark, the extended gauge groups are
generally flavor non-universal, which predict the existence of the
non-universal gauge bosons. For instance, the topcolor-assisted
technicolor ($TC2$) model [4] and the flavor universal $TC2$ model
[5] predict the non-universal neutral gauge boson $Z'$, which treats
the third generation fermions differently from those in the first
and second generations. Possible signals of the gauge boson $Z'$
have been studied in the literature [6].

The top quark with a mass of the order of the electronweak scale,
$m_{t}=171.4\pm1.2(stat.)\pm 1.8(syst.)GeV$ [7], is the heaviest
particle yet discovered, which is singled out to play a key role in
probing the new physics beyond the $SM$. Very large mass of top
quark prompts itself to rapidly decay before hadronization and the
spin information of the top quark is preserved from its decay. Thus,
the top spin configuration can provide additional observables for
testing the $SM$ and new physics beyond the $SM$. Spin correlation
in top production and decay is a very interesting issue in top quark
physics.

It is well known that, at the high energy linear $e^{-}e^{+}$
collider ($ILC$) experiments, the top quark pair ($t\bar{t}$) is
produced in a unique spin configuration and the decay products of
polarized top quark are strongly correlated to the spin axis. Thus,
studying the $t\bar{t}$ spin correlations provides an interesting
possibility to test the $SM$ and search for new physics. This fact
has made that the spin correlation of the $t\bar{t}$ production at
the $ILC$ experiments has been extensively studied in the $SM$ and
new physics beyond the $SM$ [8,9,10]. This paper is aimed at
investigating the effects of the non-universal neutral gauge boson
$Z'$ predicted by the topcolor scenario on the top quark pair
production via the process $e^{-}e^{+}\rightarrow t\bar{t}$, and
seeing whether the possible signatures of $Z'$ can be detected via
measuring the $t\bar{t}$ spin correlation observable in the future
$ILC$ experiments.

The $TC2$ model is one of the phenomenologically viable models,
which has almost all essential features of the topcolor scenario.
So, in the rest of this paper, we will give our numerical results in
detail in the context of the $TC2$ model. In the next section, we
will give the calculation formula including the contributions of the
non-universal gauge boson $Z'$ in the general spin basis. The
corrections of  $Z'$ to the $t\bar{t}$ production cross sections for
different spin configurations in the off-diagonal basis are
calculated in section III. The contributions of $Z'$ to the
$t\bar{t}$ spin correlation coefficient $C$ is discussed in section
IV. Our conclusions are given in section V.

\noindent{\bf II.The relevant formula of the  $t\bar{t}$
 production  in the general spin basis }

As it is well known, a large part of the top quark mass is
dynamically generated by topcolor interaction at a scale of order 1
$TeV$ in the $TC2$ model. To ensure that the top quark condenses and
receives a large mass while the bottom quark does not, a
non-universal extended hypercharge group $U_{y}(1)$ is introduced,
so the $TC2$ model predicts the existence of the extra $U_{y}(1)$
gauge boson $Z'$, which couples preferentially to the third
generation fermions. Thus, the new particle $Z'$ should be easily
detected via the processes involving the third generation fermions.
The couplings of $Z'$ to ordinary fermions which are related our
calculation  can be written as [3,11]:

\begin{equation}
\frac{1}{2}g_{1}[\frac{1}{3}cot\theta'Z_{\mu}'(\bar{t_{L}}\gamma^{\mu}
t_{L}+2\bar{t_{R}}\gamma^{\mu}t_{R})
+tan\theta'Z_{\mu}'(\bar{e_{L}}\gamma^{\mu}e_{L}+2\bar{e_{R}}
\gamma^{\mu}e_{R})],
\end{equation}
where $g_{1}$ is the ordinary hypercharge gauge coupling constant,
$\theta'$ is mixing angle with $tan\theta'=\frac{g_{1}}{\sqrt{4\pi
k_{1}}}$. To obtain the top quark condensation without forming a
$b\bar{b}$ condensation, there must be $k_{1}\leq 1$ [5]. In our
numerical estimation, we will assume that the value of $k_{1}$ is in
the range of $0.1\sim0.9$.

Top quark pairs will be abundantly produced via the s-channel photon
exchange and $Z$ exchange at the $ILC$ experiments. The gauge boson
$Z'$ predicted by topcolor scenario can give additional
contributions to top quark pair production. In this paper, we will
discuss $Z'$ effects on the spin configurations of the top quark
pair production in the general spin basis. In this basis, the spin
states of the top quark and top anti-quark are defined in their own
rest-frame by decomposing their spins along reference axes $\bar{A}$
and $\bar{\hat{A}}$. The reference axis $\bar{A}$ for top quark is
expressed by an angle $\xi$ between the axis and the top anti-quark
momentum in the rest frame of the top quark. The usual helicity
basis can be obtained by taking $\xi=\pi$.

In the general spin basis, the differential cross sections of the
process $e^{-}e^{+}\rightarrow t\bar{t}$ for different spin
configurations of the top quark pair
$(\uparrow\uparrow,\downarrow\downarrow,\uparrow\downarrow,\downarrow\uparrow)$
can be written as:
\begin{eqnarray}
\frac{d\sigma_{L}}{dcos\theta}(e_{L}^{-}e^{+}_{R}\rightarrow
t\uparrow \bar{t}\uparrow or \hspace{0.1cm} t\downarrow
\bar{t}\downarrow)&=&\frac{3\pi\alpha_{e}^{2}\beta}{8S}|[\sqrt{1-\beta^{2}}
sin\theta cos\xi_{L}-sin\xi_{L}(cos\theta+\beta)]g_{LL}\nonumber\\
&+&[\sqrt{1-\beta^{2}}sin\theta
cos\xi_{L}-sin\xi_{L}(cos\theta-\beta)]g_{LR}|^{2},\\
\frac{d\sigma_{L}}{dcos\theta}(e_{L}^{-}e^{+}_{R}\rightarrow
t\uparrow \bar{t}\downarrow ( t\downarrow
\bar{t}\uparrow))&=&\frac{3\pi\alpha_{e}^{2}\beta}{8S}|[\sqrt{1-\beta^{2}}
sin\theta sin\xi_{L}+cos\xi_{L}(cos\theta+\beta)\nonumber\\
&\pm&(1+\beta cos\theta)]g_{LL}+[\sqrt{1-\beta^{2}}sin\theta
sin\xi_{L}\nonumber\\
&+&cos\xi_{L}(cos\theta-\beta)\pm(1-\beta cos\theta)]g_{LR}|^{2},\\
\frac{d\sigma_{R}}{dcos\theta}(e_{R}^{-}e^{+}_{L}\rightarrow
t\uparrow \bar{t}\uparrow or \hspace{0.1cm} t\downarrow
\bar{t}\downarrow)&=&\frac{3\pi\alpha_{e}^{2}\beta}{8S}|[\sqrt{1-\beta^{2}}
sin\theta cos\xi_{R}-sin\xi_{R}(cos\theta+\beta)]g_{RR}\nonumber\\
&+&[\sqrt{1-\beta^{2}}sin\theta
cos\xi_{R}-sin\xi_{R}(cos\theta-\beta)]g_{RL}|^{2},\\
\frac{d\sigma_{R}}{dcos\theta}(e_{R}^{-}e^{+}_{L}\rightarrow
t\uparrow \bar{t}\downarrow ( t\downarrow
\bar{t}\uparrow))&=&\frac{3\pi\alpha_{e}^{2}\beta}{8S}|[\sqrt{1-\beta^{2}}
sin\theta sin\xi_{R}+cos\xi_{R}(cos\theta+\beta)\nonumber\\
&\mp&(1+\beta cos\theta)]g_{RR}+[\sqrt{1-\beta^{2}}sin\theta
sin\xi_{R}\nonumber\\
&+&cos\xi_{R}(cos\theta-\beta)\mp(1-\beta cos\theta)]g_{RL}|^{2}
\end{eqnarray}
with
\begin{eqnarray}
g_{IJ}&=&g^{SM}_{IJ}+g_{I}^{Z'}(e)g_{J}^{Z'}(t)\frac{S}{S-M_{Z'}^{2}+iM_{Z'}
\Gamma_{Z'}}\nonumber\\
&=&g_{\gamma}(e)g_{\gamma}(t)+g_{I}^{Z}(e)g_{J}^{Z}(t)\frac{S}{S-M_{Z}^{2}+iM_{Z}
\Gamma_{Z}}\nonumber\\
&+&g_{I}^{Z'}(e)g_{J}^{Z'}(t)\frac{S}{S-M_{Z'}^{2}+iM_{Z'}\Gamma_{Z'}}.
\end{eqnarray}
Where $\theta$ is the scattering angle of the top quark with respect
to the electron beam, $t\uparrow$ ($t\downarrow$) denotes the top
quark spin along (against) the reference axis $\bar{A}$, and
$\beta=\sqrt{1-\frac{4m_{t}^{2}}{S}}$. The angles $\xi_{L}$ and
$\xi_{R}$ correspond to the left- and right- handed electron beam,
respectively. $\sqrt{S}$ is the center-of-mass (c.m.) of the $ILC$
experiments. $\Gamma_{Z}$ and $\Gamma_{Z'}$ are the decay widths of
the gauge bosons $Z$ and $Z'$, respectively. $g_{I}^{Z}(e)$ and
$g_{J}^{Z}(t)$ ($g_{I}^{Z'}(e)$ and $g_{J}^{Z'}(t)$) with $I,J=L$ or
$R$ represent the couplings of the $SM$ gauge boson $Z$ (the new
gauge boson $Z'$ ) to the electron and top quark, respectively. They
can be written as:
\begin{eqnarray}
g_{\gamma}(e)=-1, \hspace{6cm} g_{\gamma}(t)&=&\frac{2}{3};\\
g_{L}^{Z}(e)=\frac{S_{W}}{C_{W}}-\frac{1}{2S_{W}C_{W}},
\hspace{3.6cm} g_{R}^{Z}(e)&=&\frac{S_{W}}{C_{W}};\\
g_{L}^{Z}(t)=\frac{1}{2S_{W}C_{W}}-\frac{2S_{W}}{3C_{W}},
\hspace{3.4cm} g_{R}^{Z}(t)&=&-\frac{2S_{W}}{3C_{W}};\\
g_{L}^{Z'}(e)=\frac{1}{4C_{W}^{2}\sqrt{\pi k_{1}}},
\hspace{4.3cm} g_{R}^{Z'}(e)&=&\frac{1}{2C_{W}^{2}\sqrt{\pi k_{1}}};\\
g_{L}^{Z'}(t)=\frac{\sqrt{\pi k_{1}}}{3}, \hspace{5.2cm}
g_{R}^{Z'}(t)&=&\frac{2\sqrt{\pi k_{1}}}{3}.
\end{eqnarray}
Where $S_{W}=sin\theta_{W}$, $\theta_{W}$ is the Weinberg angle.

The lower limits on the mass $M_{Z'}$ of the new gauge boson $Z'$
predicted by topcolor scenario can be obtained via studying its
effects on various observable, which has been extensively  studied
[3]. For example, Ref.[12] has shown that, to fit the electroweak
precision measurement data, the $Z'$ mass $M_{Z'}$ must be larger
than 1 $TeV$. The lower bounds on $M_{Z'}$ can also be obtained from
dijet and dilepton production at the Tevatron experiments [13], or
from $B\bar{B}$ mixing [14]. However, these bounds are significantly
weaker than those from precisely electroweak data. Furthermore,
Refs.[5,15] have shown that, for the coupling parameter $k_{1}<1$,
the $Z'$ mass $M_{Z'}$ can be explored up to several $TeV$ at the
$ILC$ experiment with $\sqrt{S}=500GeV$ and the integrated
luminosity $\pounds_{int}=100fb^{-1}$. As numerical estimation, we
will take $M_{Z'}$ as a free parameter and assume that $M_{Z'}$ is
in the range of $1TeV\sim 2TeV$ throughout this paper. In this case,
the total decay width $\Gamma_{Z'}$ is dominated by $t\bar{t}$ and
$b\bar{b}$ modes, which can be approximately written as [16]:
\begin{equation}
\Gamma_{Z'}\simeq\frac{g_{1}cot^{2}\theta'}{12\pi}M_{Z'}=\frac{1}{3}k_{1}M_{Z'}
\end{equation}

In the following sections, we will use above equations to discuss
the contributions of $Z'$ to different spin configurations and the
spin correlation observable for the $t\bar{t}$ production at the
$ILC$ experiments.

\noindent{\bf III. Contributions of $Z'$ to the cross sections of
$t\bar{t}$ production for different spin  \hspace*{0.8cm}
configurations }

\begin{figure}[htb]
\vspace{-0.5cm}
\begin{center}
\epsfig{file=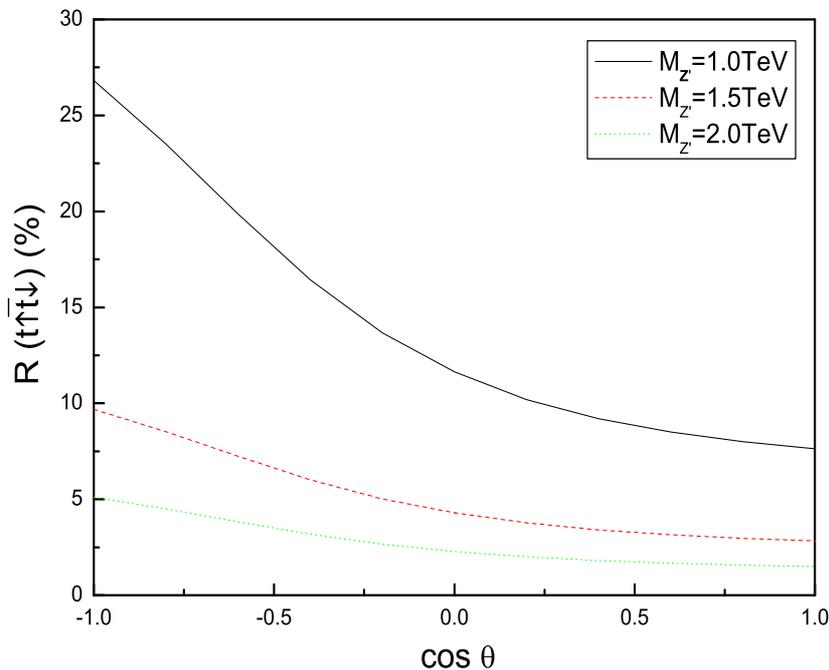,width=350pt,height=300pt} \vspace{-1cm}
\hspace{0.5cm} \caption{The relative correction parameter
$R(t\uparrow\bar{t}\downarrow)$ as a function of $cos\theta$ for
 \hspace*{1.8cm}  $\sqrt{S}=500GeV$, $k_{1}=0.6$, and three values
of $M_{Z'}$.} \label{ee}
\end{center}
\end{figure}

It has been shown that, in the $SM$, the special values of the
angles $\xi_{L}$ and $\xi_{R}$ can make the production cross
sections of the like-spin states $t\uparrow \bar{t}\uparrow$ and
$t\downarrow \bar{t}\downarrow$ vanish for the left- and right-
handed electron beam, respectively [17]. Since the production cross
sections for off-diagonal spin states ($t\uparrow \bar{t}\downarrow$
and $ t\downarrow\bar{t}\uparrow$) are non-zero, thus it is called
the off-diagonal basis. From Eq.(2) and Eq.(4), we can see that, in
the $SM$, the expressions of the angles $\xi_{L}$ and $\xi_{R}$
corresponding the off-diagonal basis are taken to be:
\begin{equation}
cos\xi_{I}=\frac{-A_{I}}{\sqrt{A^{2}_{I}+B^{2}_{I}}},
\hspace{3.2cm}
sin\xi_{I}=\frac{-B_{I}}{\sqrt{A^{2}_{I}+B^{2}_{I}}}
\end{equation}
with
\begin{equation}
A_{I}=g^{SM}_{II}(cos\theta+\beta)+g^{SM}_{IJ}(cos\theta-\beta),
\end{equation}
\begin{equation}
B_{I}=(g^{SM}_{II}+g^{SM}_{IJ})sin\theta\sqrt{1-\beta^{2}}.
\end{equation}

Obviously, there are two off-diagonal bases for $t\bar{t}$
production, one is for $e^{-}_{L}e_{R}^{+}$ scattering and the other
is for $e^{-}_{R}e_{L}^{+}$. However, for the process
$e^{-}e^{+}\rightarrow t\bar{t}$, the two spin bases are almost
identical. Thus, we will use the off-diagonal basis for
$e^{-}_{L}e_{R}^{+}$ scattering even when discussing
$e^{-}_{R}e_{L}^{+}$ scattering in our numerical calculation.

\begin{figure}[htb]
\vspace{-0.5cm}
\begin{center}
\epsfig{file=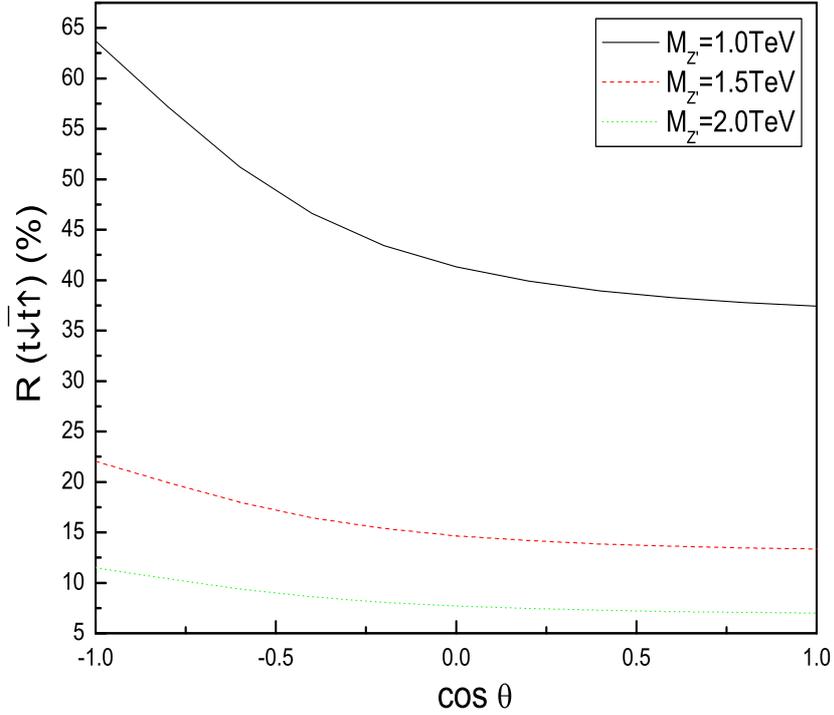,width=350pt,height=320pt} \vspace{-1cm}
\hspace{0.5cm} \caption{Same as Fig.1 but for the relative
correction parameter $R(t\downarrow\bar{t}\uparrow)$ .} \label{ee}
\end{center}
\end{figure}

One interesting feature of the off-diagonal basis is that, in the
$SM$, the production cross sections
$\sigma_{L}(t\uparrow\bar{t}\downarrow)$ and
$\sigma_{R}(t\downarrow\bar{t}\uparrow)$ are dominant for the
production cross sections $\sigma_{L}$ of the process
$e_{L}^{-}e_{R}^{+}\rightarrow t\bar{t}$ and $\sigma_{R}$ of the
process $e_{R}^{-}e_{L}^{+}\rightarrow t\bar{t}$, respectively. The
differential cross section is an observable containing more detailed
information about the underlying dynamics of the relevant process.
To see the effects of the non-universal gauge boson $Z'$ on the
dominant production modes, we define the relative correction
parameters as:
\begin{equation}
R(t\uparrow\bar{t}\downarrow)=\frac{\frac{d\sigma_{L}^{SM+Z'}(e_{L}^{-}e_{R}^{+}\rightarrow
t\uparrow\bar{t}\downarrow)}{dcos\theta}-\frac{d\sigma_{L}^{SM}(e_{L}^{-}e_{R}^{+}\rightarrow
t\uparrow\bar{t}\downarrow)}{dcos\theta}}{\frac{d\sigma_{L}^{SM}(e_{L}^{-}e_{R}^{+}\rightarrow
t\uparrow\bar{t}\downarrow)}{dcos\theta}}
\end{equation}
and
\begin{equation}
R(t\downarrow\bar{t}\uparrow)=\frac{\frac{d\sigma_{R}^{SM+Z'}(e_{R}^{-}e_{L}^{+}\rightarrow
t\downarrow\bar{t}\uparrow)}{dcos\theta}-\frac{d\sigma_{R}^{SM}(e_{R}^{-}e_{L}^{+}\rightarrow
t\downarrow\bar{t}\uparrow)}{dcos\theta}}{\frac{d\sigma_{R}^{SM}(e_{R}^{-}e_{L}^{+}\rightarrow
t\downarrow\bar{t}\uparrow)}{dcos\theta}}.
\end{equation}

\begin{figure}[htb]
\vspace{-0.5cm}
\begin{center}
\epsfig{file=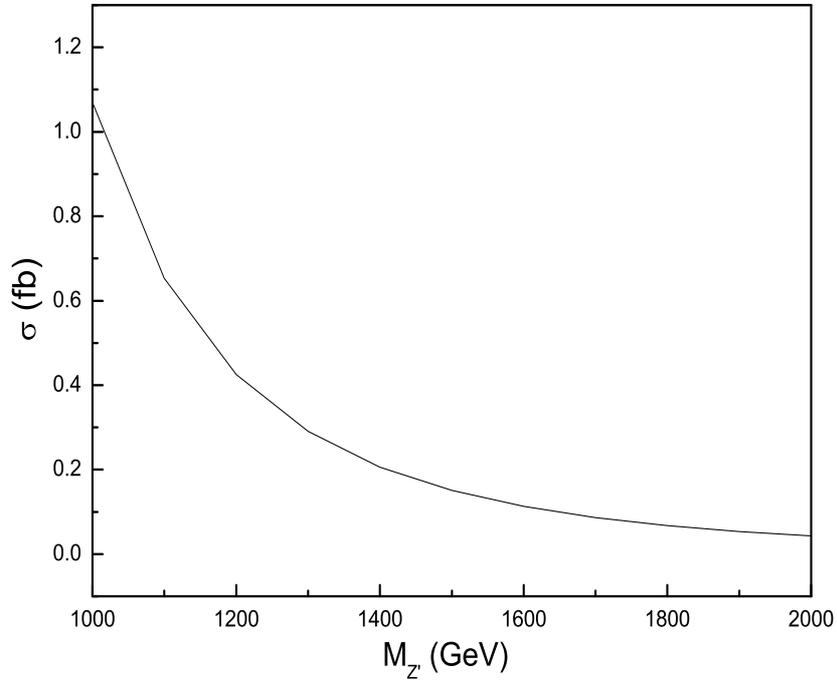,width=350pt,height=300pt} \vspace{-0.7cm}
\hspace{0.5cm} \caption{The production cross section $\sigma$ for
the like-spin states as a function of the $Z'$  \hspace*{1.8cm} mass
$M_{Z'}$ for $\sqrt{S}=500GeV$ and $k_{1}=0.6$ .} \label{ee}
\end{center}
\end{figure}

In Fig.1 and Fig.2, we plot the relative correction parameters
$R(t\uparrow\bar{t}\downarrow)$ and $R(t\downarrow\bar{t}\uparrow)$
as functions of $cos\theta$ for $\sqrt{S}=500GeV$ and three values
of the $Z'$ mass $M_{Z'}$. $R(t\uparrow\bar{t}\downarrow)$ and
$R(t\downarrow\bar{t}\uparrow)$ are dependent on the coupling
parameter $k_{1}$ only via the total decay width $\Gamma_{Z'}$.
Thus, their values are insensitive to the free parameter $k_{1}$.
 We have taken $k_{1}=0.6$ in Fig.1 and Fig.2. From these figures,
 one can see that the values of the relative correction parameters
decrease as $M_{Z'}$ increasing and $cos\theta$ increasing. For
$1TeV\leq M_{Z'}\leq 2TeV$ and $cos\theta\leq 0$, the values of
$R(t\uparrow\bar{t}\downarrow)$ and $R(t\downarrow\bar{t}\uparrow)$
are in the ranges of $27\%\sim 2.3\% $ and $64\%\sim 7.7\% $,
respectively. The contributions of $Z'$ to the process
$e_{R}^{-}e_{L}^{+}\rightarrow t\downarrow\bar{t}\uparrow$ are
larger than those for the process $e_{L}^{-}e_{R}^{+}\rightarrow
t\uparrow\bar{t}\downarrow$, which is because the right-handed
couplings of $Z'$ to fermions are larger than the left-handed
couplings of $Z'$ to fermions.

\begin{figure}[htb] \vspace{-0.5cm}
\begin{center}
\epsfig{file=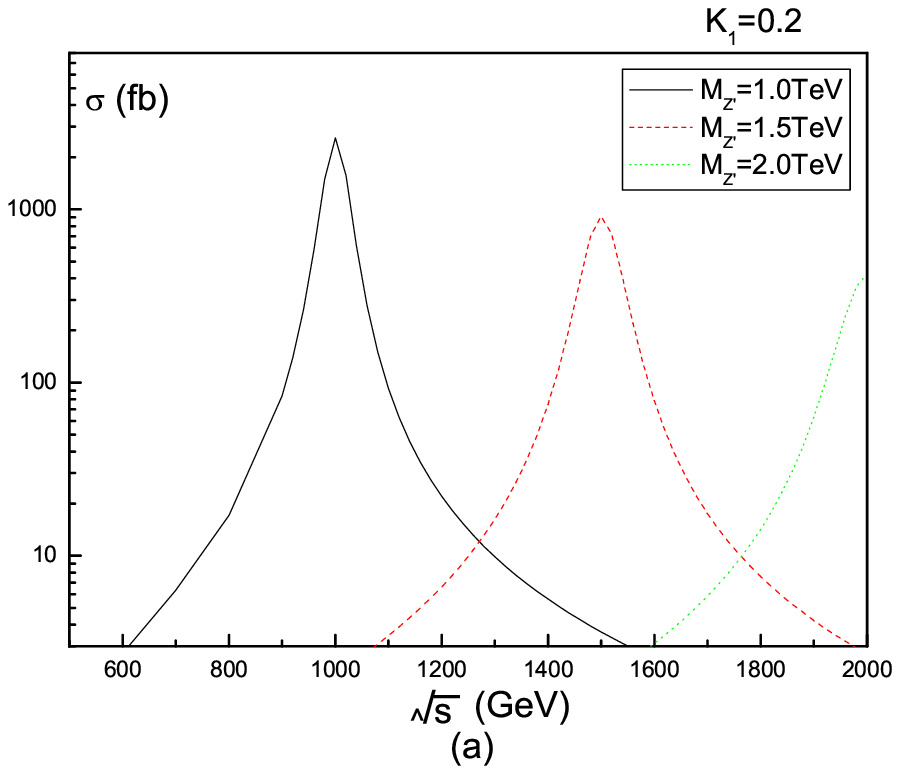,width=220pt,height=205pt}
\hspace{0cm}\vspace{-0.25cm}
\epsfig{file=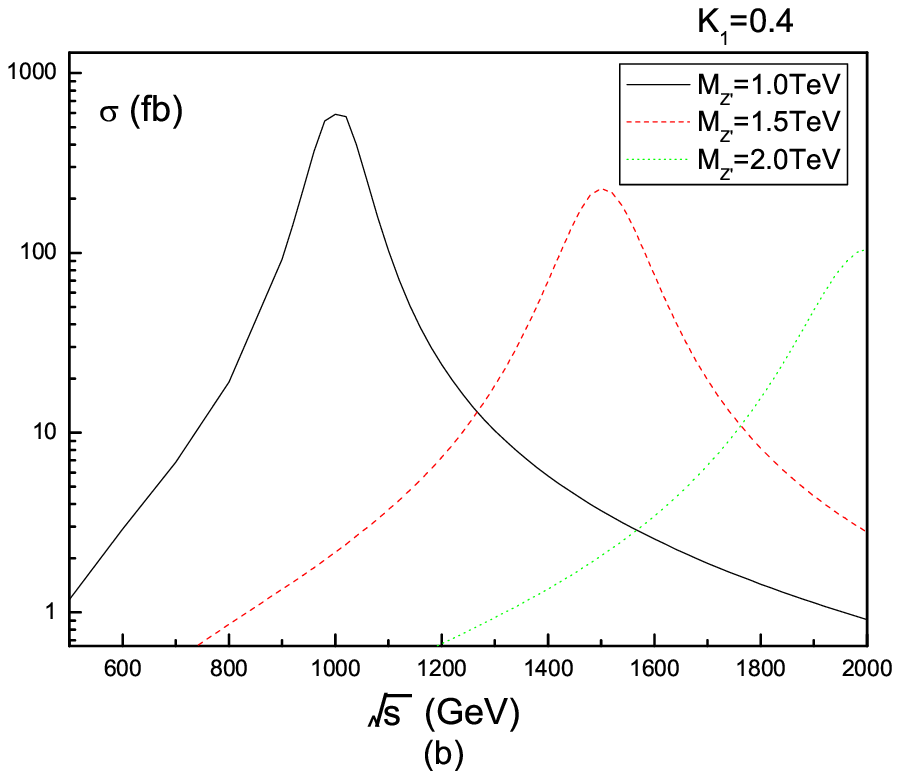,width=220pt,height=205pt} \hspace{-0.5cm}
\hspace{10cm}\vspace{-1cm}
\epsfig{file=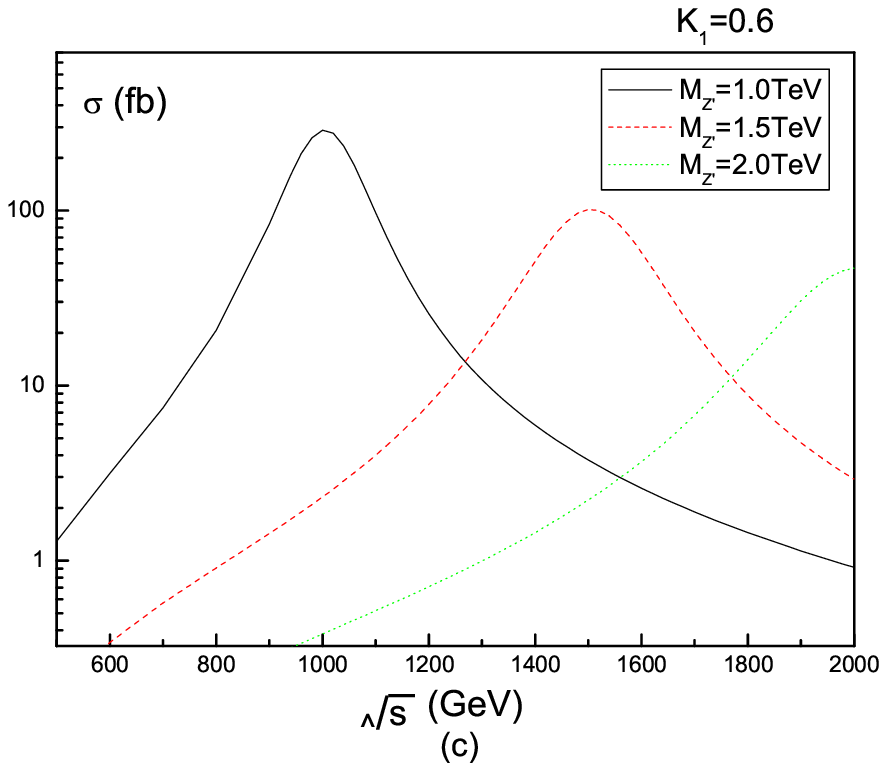,width=220pt,height=205pt} \vspace{+0.2cm}
\hspace{0cm}\vspace{-0.25cm}
\epsfig{file=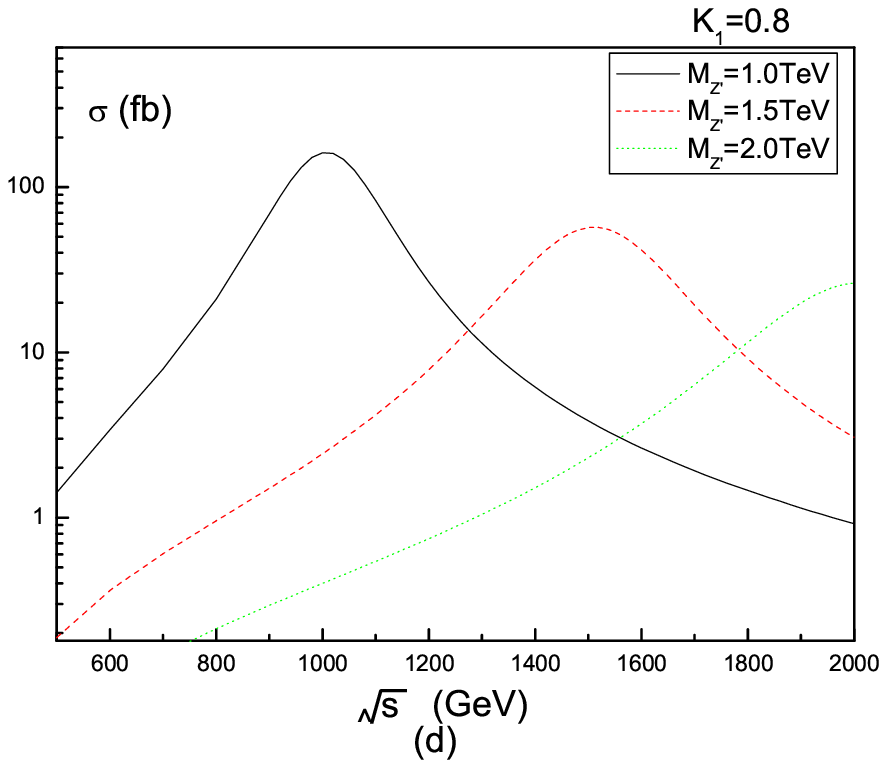,width=220pt,height=205pt} \hspace{0.5cm}
\caption{The production cross section $\sigma$ for the like-spin
states as a function of the  \hspace*{1.8cm} c.m. energy $\sqrt{S}$
for different values of $M_{Z'}$ and $k_{1}$. }\label{ee}
\end{center}
\end{figure}

In the off-diagonal basis, the  cross section of $t\bar{t}$
production predicted by the $SM$ vanish for the like-spin states
($\uparrow\uparrow,\downarrow\downarrow$) . However, $Z'$ exchange
can contribute to the relevant process
$e_{L}^{-}e_{R}^{+}+e_{R}^{-}e_{L}^{+}\rightarrow
t\uparrow\bar{t}\uparrow+t\downarrow\bar{t}\downarrow$. In Fig.3, we
plot the production cross section $\sigma$ of this process
 as a function
of $M_{Z'}$ for $\sqrt{S}=500GeV$ and $k_{1}=0.6$. One can see from
Fig.3 that, for $1TeV\leq M_{Z'}\leq 2TeV$, the value of $\sigma$ is
in the range of $1.07fb\sim 4.2\times 10^{-2}fb$. If we assume that
the yearly integrated luminosity of the $ILC$ experiment with
$\sqrt{S}=500GeV$ is $\pounds_{int}=340fb^{-1}$[18], then there will
be several tens of the raw events of the like-spin top quark pairs
to be generated per year. Certainly, the degrees of electron and
positron polarization and the detecting efficiency for the final
state particles are generally smaller than 100\%, the number of
observable events is smaller than the number of the raw events.

To see the effects of the c.m. energy $\sqrt{S}$ on the production
cross section $\sigma$ for the like-spin states, $\sigma$ is shown
in Fig.4 as a function of $\sqrt{S}$ for different values of
$M_{Z'}$ and $k_{1}$. One can see from Fig.4 that the cross section
resonance emerges when  $M_{Z'}$ approaches the c.m. energy
$\sqrt{S}$.  The resonance values of the cross section $\sigma$ are
strongly dependent on the $Z'$ mass $M_{Z'}$ and the coupling
parameter $k_{1}$, which decrease as $M_{Z'}$ increasing and $k_{1}$
increasing. For $M_{Z'}=1.5TeV,k_{1}=0.4$ and $0.8$, the maximum
values of the production cross section $\sigma$ can reach $228fb$
and $57fb$, respectively. Thus, the resonance effects of the gauge
boson $Z'$ on the process $e^{-}e^{+}\rightarrow
t\uparrow\bar{t}\uparrow+t\downarrow\bar{t}\downarrow$ might be
observed in the future $ILC$ experiments.

Of course, in practice, there are experimental issues to consider,
that will dilute the signal. We must take into account the fact that
the colliding electron and positron will not have $ 100\% $
polarization. Moreover, there will also be a considerable dilution
of the signal associated with the efficiency of measuring the
polarization of the top quarks.

\noindent{\bf IV. The non-universal gauge boson $Z'$ and the
$t\bar{t}$ spin correlation observable }

From above discussions we can see that, at the $ILC$, the top quark
pair $t\bar{t}$ can be produced in an unique spin configuration. In
the familiar helicity basis, all spin configurations have
contributions to the production cross section of the process
$e^{-}e^{+}\rightarrow t\bar{t}$. In the off-diagonal basis, the
contribution from the like-spin states $(t\uparrow \bar{t}\uparrow$
and $t\downarrow \bar{t}\downarrow)$ to the $SM$ process
$e^{-}e^{+}\rightarrow t\bar{t}$ at the tree-level is zero, while
the contributions from the spin states $t\uparrow \bar{t}\downarrow$
and $t\downarrow\bar{t}\uparrow$ are dominant for the cross sections
of the processes $e^{-}_{L}e^{+}_{R}\rightarrow t\bar{t}$ and
$e^{-}_{R}e^{+}_{L}\rightarrow t\bar{t}$, respectively.
\begin{figure}[htb] \vspace{-0.5cm}
\begin{center}
\epsfig{file=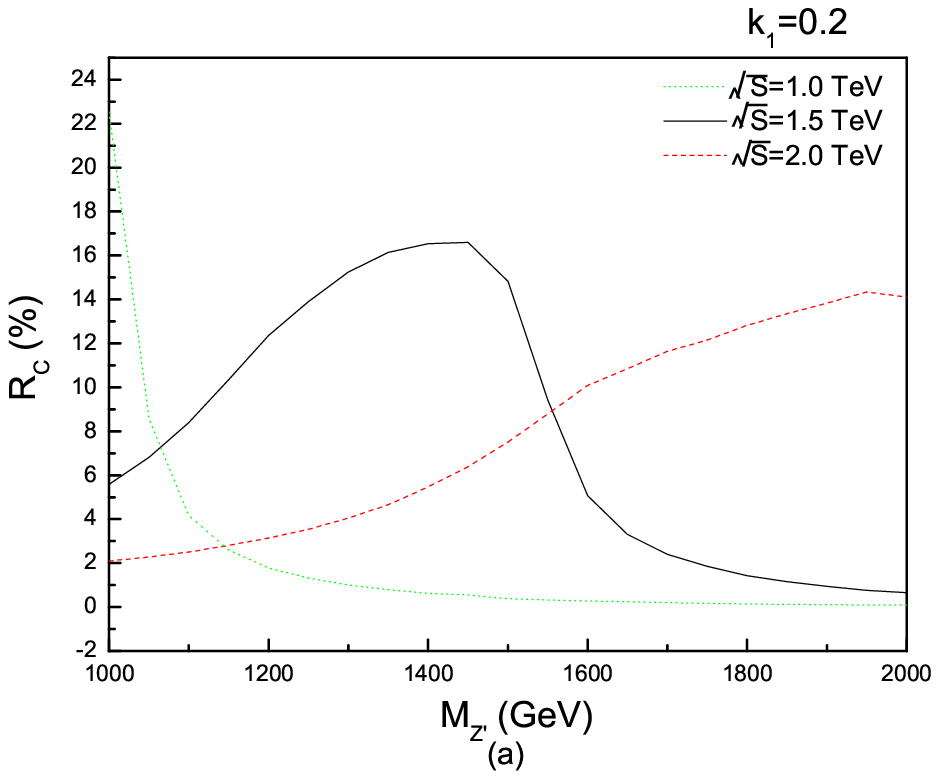,width=220pt,height=205pt}
\hspace{0cm}\vspace{-0.25cm}
\epsfig{file=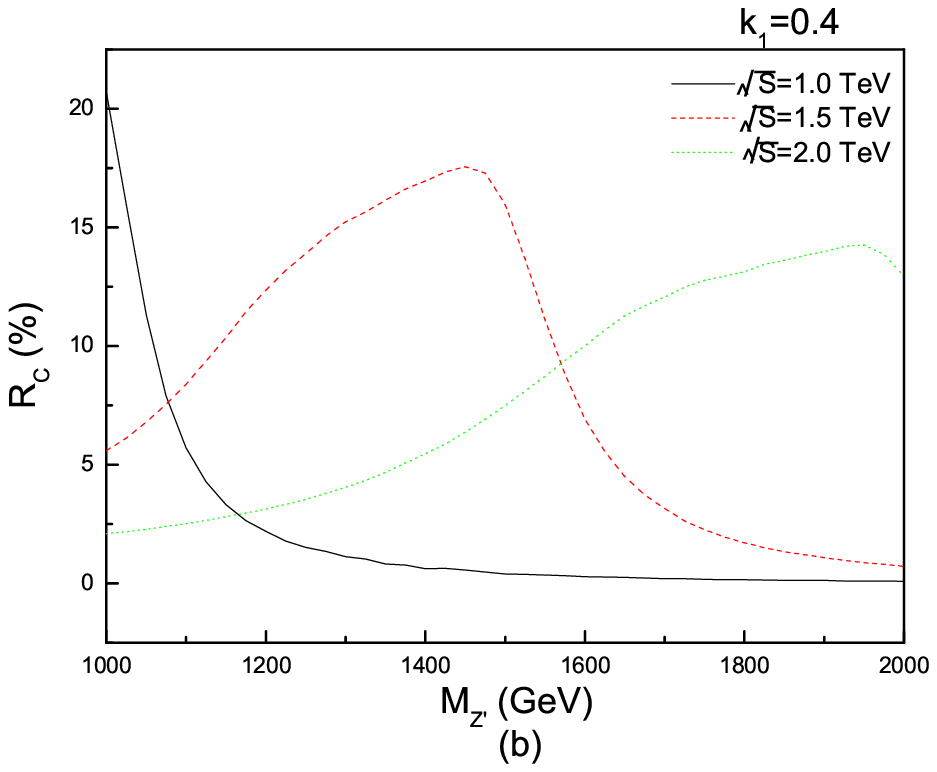,width=220pt,height=205pt} \hspace{-0.5cm}
\hspace{10cm}\vspace{-1cm}
\epsfig{file=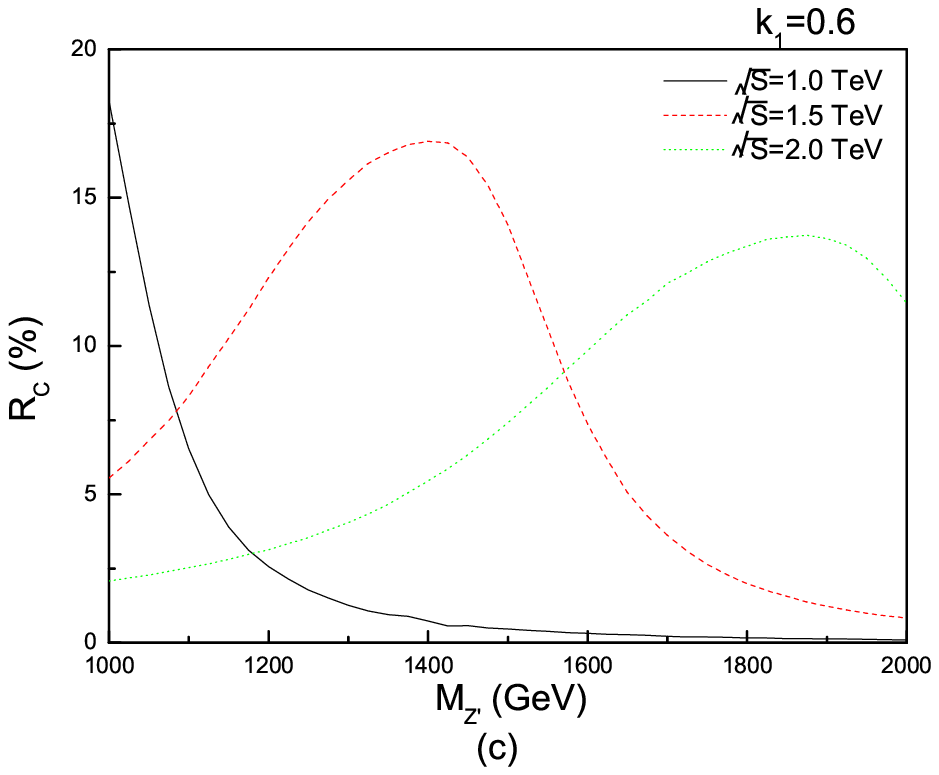,width=220pt,height=205pt} \vspace{+0.2cm}
\hspace{0cm}\vspace{-0.25cm}
\epsfig{file=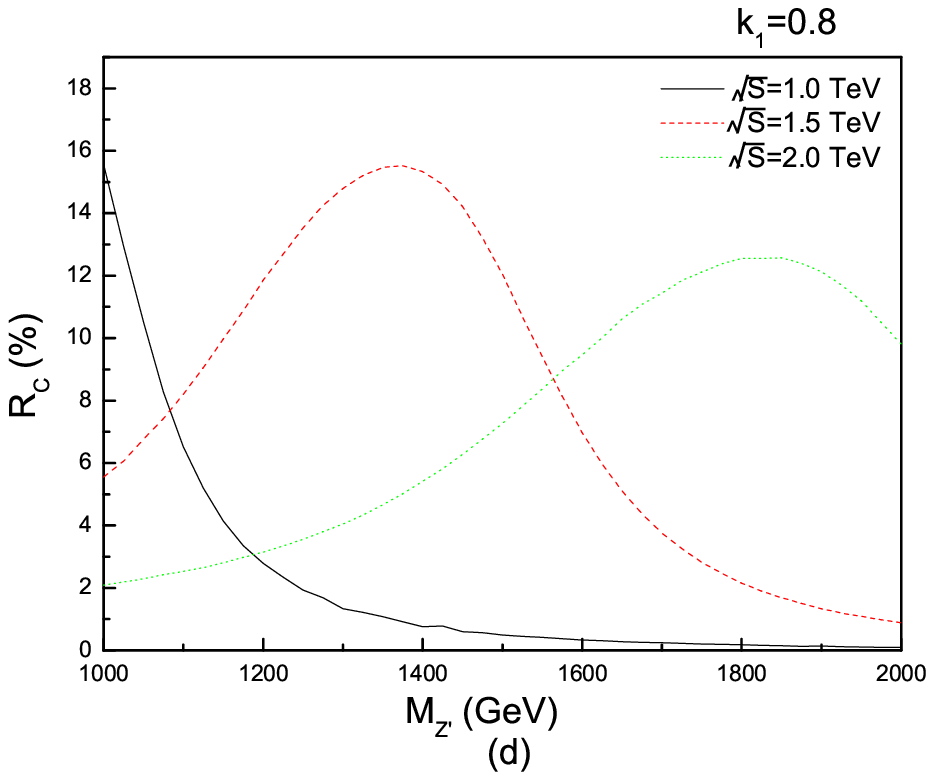,width=220pt,height=205pt} \hspace{0.5cm}
\caption{The relative correction parameter
$R_{c}=\frac{\delta_{C}}{C^{SM}}$as a function of the $Z'$ mass
$M_{Z'}$ \hspace*{1.8cm} for different values of $\sqrt{S}$ and
$k_{1}$.}\label{ee}
\end{center}
\end{figure}
This means that, in the off-diagonal basis, all $t\bar{t}$ pairs
predicted by the $SM$ at tree-level are of opposite spins, the decay
products of polarized top quark are strongly correlated to the spin
axis, the top quark events at the $ILC$ have a very distinctive
topology. However, the non-universal gauge boson $Z'$ predicted by
topcolor scenario has contributions to the like-spin states
$t\uparrow \bar{t}\uparrow$ and $t\downarrow \bar{t}\downarrow$,
which can change this topology. Thus, the gauge boson $Z'$ can
product corrections to the $t\bar{t}$ spin correlation coefficient
$C$.

The $t\bar{t}$ spin correlation can be measured by analyzing the
angular distributions of the $t$ and $\bar{t}$ decay products. In
principle, the process $e^{-}e^{+}\rightarrow t\bar{t} \rightarrow
6$ jets is very complex. However, since $\Gamma_{t}\ll m_{t}$, the
narrow width approximation approach is valid for $t$ and $\bar{t}$
quarks. Using this method, it has been shown that the best way to
analyze the $t\bar{t}$ correlation is through angular correlation
among the two charged leptons $l$ and $l'$ in the di-lepton final
state [19]. To obtain the differential cross section of the process
$e^{-}e^{+}\rightarrow t\bar{t}\rightarrow
b\bar{l}\nu_{l}\bar{b}l'\nu_{l'}$, the production and decay spin
density matrixes are written as the usual approach. After
integrating over the azimuthal angles of the charged leptons, one
can obtain the following double differential distribution [9]
\begin{equation}
\frac{1}{\sigma}\frac{d^{2}\sigma}{dcos\theta_{\bar{l}}dcos\theta_{l'}}=\frac{1}{4}
(1+B_{1}cos\theta_{\bar{l}}+B_{2}cos\theta_{l'}-C
cos\theta_{\bar{l}}cos\theta_{l'}),
\end{equation}
where $\theta_{\bar{l}}(\theta_{l'})$ denotes the angle of the
charged lepton $\bar{l}$  $(l')$ with respect to the chosen spin
axis in the top quark (top antiquark) rest frame. The coefficients
$B_{1}(B_{2})$ and $C$ are related to the mean $t(\bar{t})$
polarization and spin correlation projected into the direction of
the spin axis. The spin correlation coefficient $C$ can be written
as:
\begin{equation}
C=\alpha_{\bar{l}}\alpha_{l'}\langle s_{t}s_{\bar{t}}\rangle,
\end{equation}
where $\alpha_i$ is the spin analyzing power of particle $i$, which
have opposite signs for the charged leptons $\bar{l}$ and $l$. At
the leading order, there are $\alpha_{\bar{l}}=-\alpha_{l}=-1$.

The contributions of the non-universal gauge boson $Z'$ to the spin
correlation coefficient $C$ can be obtained by using the formula
given in Sec.II. Our numerical results are given in Fig.5, in which
we plot the relative correction parameter
$R_{c}=\frac{\delta_{C}}{C^{SM}}$ as a function of the c.m. energy
$\sqrt{S}$ for different values of the $Z'$ mass $M_{Z'}$ and the
coupling parameter $k_{1}$.  From these figures one can see that,
when the value of the $Z'$ mass $M_{Z'}$ approaches the c.m. energy
$\sqrt{S}$ , the gauge boson $Z'$ can generate significantly
contributions to the observable $C$. For example, for $k_{1}=0.6$
and $M_{Z'}\simeq\sqrt{S}=1.5TeV$ the value of the relative
correction parameter $R_{c}$ can reach 12.5\%.  Obviously, when the
$Z'$ mass $M_{Z'}$ is much smaller or larger than the c.m. energy
$\sqrt{S}$, the effects of the non-universal gauge boson $Z'$ on
 the spin correlation coefficient $C$ are very small, which can not
 be detected at the $ILC$ experiments.

\noindent{\bf V. Conclusions }

Top quark decays before it hadronizes and the spin information of
the top quark is not spoiled and directly reflect to the
distributions of its decay products. Hence, we can utilize the top
quark spin for disentangling different top quark interactions
efficiently. The spin correlation in top quark production and
decays is an interesting issue in top quark physics.

The top quark pairs can be produced at large rates via the process
$e^{-}e^{+}\rightarrow t\bar{t}$ in a clear environment at the
$ILC$, which can be seen as an ideal tool to make precision
measurement of the top quark properties. In the off-diagonal basis,
the process $e^{-}e^{+}\rightarrow t\bar{t}$ induced by the $SM$ has
two characteristic features. One is that the production cross
sections for the like-spin states vanish and the other is that the
production cross section of the spin state
$t\uparrow\bar{t}\downarrow (t\downarrow\bar{t}\uparrow$) for the
left-handed (right-handed) electron beam is dominant. The effects of
the $QCD$ corrections do not change these features. Thus,
observation of sizable cross section for the like-spin states and
large corrections
 to the cross sections for the
$t\uparrow\bar{t}\downarrow$ and $t\downarrow\bar{t}\uparrow$ states
can be seen as signals of new physics beyond the $SM$.

In this paper, we calculate the contributions of the non-universal
gauge boson $Z'$ predicted by the $TC2$ model to the spin
configurations of the top quark pair production via the process
$e^{-}e^{+}\rightarrow t\bar{t}$ in the off-diagonal basis. Our
numerical results show that $Z'$ exchange can generate significantly
corrections to the differential cross sections for the
$t\uparrow\bar{t}\downarrow$ and $t\downarrow\bar{t}\uparrow$
states. For $1TeV\leq M_{Z'}\leq 2TeV$ and $cos\theta\leq 0$, the
values of the relative correction parameters
$R(t\uparrow\bar{t}\downarrow)$ and $R(t\downarrow\bar{t}\uparrow)$
are in the ranges of $26\%\sim 2.3\% $ and $64\%\sim 7.7\% $,
respectively. The production cross section for the like-spin state
$t\uparrow\bar{t}\uparrow+t\downarrow\bar{t}\downarrow$ is non zero
induced by $Z'$ exchange, which is in the range of $1.07fb\sim
4.2\times 10^{-2}fb$ for $\sqrt{S}=500GeV$, $0.2\leq k_{1}\leq 0.8$,
and $1TeV\leq M_{Z'}\leq 2TeV$. Furthermore, when the $Z'$ mass
$M_{Z'}$ approaches the c.m. energy $\sqrt{S}$, the cross section
resonance emerges. For $k_{1}=0.2$, $M_{Z'}\simeq \sqrt{S}=1TeV $
and $1.5TeV$, the resonance values of the cross section
$\sigma(t\uparrow\bar{t}\uparrow+t\downarrow\bar{t}\downarrow)$ can
reach $2586fb$ and $912.5fb$ , respectively.

To see whether the effects of $Z'$ on the $t\bar{t}$ production can
be measured in the future $ILC$ experiments, we further calculate
the contributions of $Z'$ to the $t\bar{t}$ spin correlation
coefficient $C$. We find that, with reasonable values of the free
parameters $M_{Z'}$ and $k_{1}$, the value of the relative
correction parameter $R_{c}$ can be significantly large. Thus, we
expect that the effects of the new gauge boson $Z'$ on the spin
configurations of the $t\bar{t}$ production and the $t\bar{t}$ spin
correlation coefficient $C$ might be observed at the future $ILC$
experiments.

\vspace{0.5cm} \noindent{\bf Acknowledgments}

This work was supported in part by Program for New Century Excellent
Talents in University(NCET-04-0290), the National Natural Science
Foundation of China under the Grants No.10475037 and 10675057.

\end{document}